\documentclass[preprint,aps]{revtex4}

\usepackage{graphicx}
\usepackage{multirow}
\usepackage{makecell}
\usepackage{booktabs}

\begin{document}

\title{ Detailed Electronic Structure of the Three-Dimensional Fermi Surface and its Sensitivity to Charge Density Wave Transition in ZrTe$_3$ Revealed by High Resolution Laser-Based Angle-Resolved Photoemission Spectroscopy}

\author{Shou-Peng Lyu$^{1,2}$, Li Yu$^{1,*}$, Jian-Wei Huang$^{1,2}$, Cheng-Tian Lin$^{3}$, Qiang Gao$^{1,2}$, Jing Liu$^{1,2}$, Guo-Dong Liu$^{1}$, Lin Zhao$^{1}$, Jie Yuan$^{1}$, Chuang-Tian Chen$^{4}$, Zu-Yan Xu$^{4}$, and Xing-Jiang Zhou$^{1,2,5,*}$}

\affiliation{
\\$^{1}$National Lab for Superconductivity, Beijing National Laboratory for Condensed Matter Physics, \\Institute of Physics, Chinese Academy of Sciences, Beijing 100190, China
\\$^{2}$University of Chinese Academy of Sciences, Beijing 100049, China
\\$^{3}$Max-Planck-Institut f$\ddot{u}$r Festk$\ddot{o}$rperforschung, Heisenbergstrasse 1, 70569 Stuttgart, Germany
\\$^{4}$Technical Institute of Physics and Chemistry, Chinese Academy of Sciences, Beijing 100190, China
\\$^{5}$Collaborative Innovation Center of Quantum Matter, Beijing 100871, China
}

\date{\today}

%
%

\begin{abstract}
The detailed information of the electronic structure is the key for understanding the nature of charge density wave (CDW) order and its relationship with superconducting order in microscopic level. In this paper, we present high resolution laser-based angle-resolved photoemission spectroscopy (ARPES) study on the three-dimensional (3D) hole-like Fermi surface around the Brillouin zone center in a prototypical qusi-one-dimensional CDW and superconducting system ZrTe$_3$. Double Fermi surface sheets are clearly resolved for the 3D hole-like Fermi surface around the zone center. The 3D Fermi surface shows a pronounced shrinking with increasing temperature.  In particular, the quasiparticle scattering rate along the 3D Fermi surface experiences an anomaly near the charge density wave transition temperature of ZrTe$_3$ ($\sim$63 K).  Signature of electron-phonon coupling is observed with a dispersion kink at $\sim$20 meV; the strength of the electron-phonon coupling around the 3D Fermi surface is rather weak. These results indicate that the 3D Fermi surface is also closely connected to the charge-density-wave transition  and  suggest a more global impact on the entire electronic structure induced by CDW phase transition in ZrTe$_3$.
\end{abstract}



\maketitle

How the superconducting phase competes or coexists with various magnetic or charge-ordering phase is a long standing fundamental issue in modern condensed matter physics. Especially in the low dimensional systems like high T$_c$ cuprates\cite{Gabovich2010}, heavy fermion superconductors\cite{Pfleiderer2009} and iron-based superconductors\cite{He2013}, superconducting order can emerge in the vicinity of multiple-order environment. More and more experimental evidences that have been collected recently point to a close relationship between superconductivity and the phase competition or phase coexistence\cite{YZhang2016XZhu,SYLi2016SCui}. This triggers, on the other hand, great interest  in various classical charge density wave (CDW) systems with  coexisting superconductivity\cite{NMiyakawa2017SJDenholme}. These systems can serve as a playground for investigating the nature of CDW order, and most importantly, its  relationship with superconductivity. Examples along this line include many kinds of systems like 1T-TiSe$_2$\cite{TCChiang2002TEKidd,MBatzill2017SKolekar},  TbTe$_3$\cite{CWKwangHua2012CWKwangHua,ZXShen2008FSchmitt}, and etc. Among all these materials, ZrTe$_3$ is one of the prototypical quasi-one-dimensional (quasi-1D) systems undergoing both a CDW transition at $\sim$63 K and a superconducting transition at $\sim$2 K\cite{STanda2012KYamaya}. Further pressure-dependent measurements\cite{YUwatoko2005RYomo,YUwatoko2017STsuchiya,STanda2002KYamaya} and ion substitution study in ZrTe$_3$\cite{YZhang2016XZhu,CPetrovic2014CMirri,CPetrovic2011XZhu,CPetrovic2011HCLei} revealed unusual connection between CDW and superconductivity, suggesting a competing-type relationship between them at low temperature. This makes ZrTe$_3$ a unique candidate to study the  complexity behind the CDW-superconductivity entanglement  in such a qusi-1D system.

To determine the detailed electronic structure and related Fermi surface and electron dynamics for better understanding the physical properties of materials in a microscopic level, ARPES is one of the most direct techniques. The semi-metallic character of ZrTe$_3$  and Fermi surface topology have been studied by several earlier ARPES measurements\cite{KYamaya2005TYokoya,HBerger2009MHoesch}. Up to now, most ARPES studies on ZrTe$_3$ concentrate on the quasi-1D Fermi surface sheet which is believed to be most relevant for the CDW formation. A  partial gap or pseudogap feature is observed around the D-point region at the Brillouin Zone (BZ) corner which was associated with strongly fluctuating CDW order that kicks in at a high temperature above 200 K\cite{KYamaya2005TYokoya}, even though the CDW transition temperature is at a much lower temperature 63 K. Such strong fluctuating character of CDW order is common in many low-dimensional systems. The nearly commensurate Fermi surface nesting vector q$_n$ deduced by connecting neighbor pseudogap regions is consistent with the CDW vector q$_{CDW}$=(1/14,0,1/3) determined from direct electron microscope measurements\cite{JAWilson1984DJEaglesham}, which supports a conventional Fermi surface nesting mechanism of CDW formation in ZrTe$_3$\cite{SLCooper2015SLGleason,MKrisch2009MHoesch}. However, recent Raman scattering measurement suggested that electron-phonon coupling can play a dominant role for the CDW order beyond the conventional nesting picture\cite{LYuan2015YWHu}. It remains to be investigated whether the rest part of the quasi-1D Fermi surface  and the 3D Fermi surface around $\Gamma$, are possibly responsible or responsive to CDW order  in ZrTe$_3$ system.  It is still an open question on how and where superconductivity emerges in the  momentum space in ZrTe$_3$.


In this paper, we present detailed ARPES measurements on the 3D Fermi surface and its associated electron dynamics by  using a newly developed laser-based ARPES system with ultra-high instrumental resolution. We resolved clearly the double-Fermi surface sheets in the 3D hole-like Fermi surface around $\Gamma$ point. The large shrinking of the 3D Fermi surface with increasing temperature is observed. In particular, the quasiparticle scattering rate along the 3D Fermi surface exhibits an anomaly near the CDW transition temperature $\sim$63 K, indicating the sensitivity of the 3D Fermi surface to the CDW transition. Signature of electron-phonon coupling is revealed with a dispersion kink at $\sim$20 meV; the electron-phonon coupling strength is rather weak. These observations provide new and comprehensive information on the Fermi surface topology and electron dynamics on the 3D Fermi surface in ZrTe$_3$ that are important for understanding CDW formation, superconductivity and their relationship in ZrTe$_3$.

The ZrTe$_3$ single crystals used in this work were prepared by chemical vapor transport method with iodine as the transport agent. As shown in the resistivity measurement of ZrTe$_3$ in Fig. 1(b), a clear hump like resistivity anomaly pops out at about 63 K along the a-axis (blue solid-line) which is attributed to the CDW formation, while a filamentary superconducting transition occurs below 2 K.  Such a CDW signature is not present in the resistivity-temperature dependence along the b-axis (red dashed line in Fig. 1(b)).  ARPES measurements were performed at our new laser-based system equipped with  6.994 eV vacuum ultraviolet laser based on non-linear optical crystal KBBF. It is equipped with the time-of-flight electron energy analyser (ARToF 10K by Scienta Omicron) which has two-dimensional probing capability in momentum space, i.e.,it can detect all photoelectrons simultaneously within a detector receiving angle 30$^{\circ}$ ($\pm$15$^{\circ}$). The energy resolution is $\sim$1meV, and the angular resolution is $\sim$0.1$^{\circ}$. Some detailed description of the ARPES system can be found in reference\cite{XJZhou2018}. Samples were cleaved $in\ situ$ at 20 K and measured in ultrahigh vacuum with a base pressure better than 5$\times$10$^{-11}$ mbar.

The crystal structure of ZrTe$_3$ (space group P21/m, Fig. 1(a)) consists of infinitely stacked ZrTe$_3$ trigonal prisms along the b-direction, forming quasi-1D prismatic chains\cite{CWChu2013XZhu,FRWagner1998KStowe}. The monoclinic unit cell contains two neighboring chains related reversely with each other and binded through nearest inter-chain Zr-Te(1) bonds to form layers in the ab-plane. The inter-layer bonding, however,  relies on Van de Waals force which makes cleaving of the single crystal sample naturally along the ab-planes. Within each layer, the nearest Te(2) and Te(3) atoms bond together to form a dimerised chain along the a-axis\cite{FRWagner1998KStowe}.  This is widely believed to be the essential element for the electronic properties in the CDW state. Band structure calculations\cite{WTremel1998CFelser} suggest that the Fermi surface of the ZrTe$_3$ consists of two major components, as sketched in Fig. 2(a), a 3D hole-like Fermi surface sheet (blue) around the BZ center which is formed by the hybridization between Zr-4d orbitals and the Te(1)-5p orbitals via nearest inter-prism hopping, and a quasi-1D electron-like Fermi surface (green) at the BZ boundary along b* direction, which is formed by the nearest hopping via Te(2)- and Te(3)-5p orbitals along the dimerised chains. Corresponding 2D projected BZ is plotted underneath the 3D BZ that are used in the ARPES measurements. The letters $\Gamma$, B, Y, and D denote the high symmetry points in the projected BZ.

Figure  2  shows the Fermi surface and constant energy contours of ZrTe$_3$ measured at 30 K. Typical bands structure along some typical momentum cuts are shown in Fig. 3.  Thanks to the  high efficiency of our new ARToF analyzer-based ARPES measurements, the entire 3D hole-like Fermi surface can be covered by two measurements, one is to cover the central Fermi surface region (Fig. 2(b)) while the other is to cover the Fermi surface tip region (Fig. 2(d)).  After careful joining these two measurement results, and symmetrize in two-fold manner within the first BZ, a complete description of the 3D hole-like Fermi surface is obtained,  as shown in Fig. 2(b). It reveals a clear Fermi surface with an anisotropic oval-shape centered at $\Gamma$ point. This Fermi surface shape is consistent with the results from earlier ARPES results\cite{KYamaya2005TYokoya,HBerger2009MHoesch}.  Fig. 2(c) and (d) present the constant energy contours of ZrTe$_3$ at different binding energies E$_b$; three characteristic energies are picked up here, E$_b$ = 0, E$_b$ = 180 meV, and E$_b$ = 300 meV, respectively. The contour area increases with increasing E$_b$, which is consistent with the hole-like nature of the 3D Fermi surface. At E$_b$ = 180 meV, there appears a shadow feature near the BZ center due to the touching  of  another hole-like bands near $\Gamma$. At E$_b$ = 300 meV, a new oval feature appears near the BZ center due to the second hole-like band near $\Gamma$, as is shown in the band structure in Fig. 3.

Very clear band splitting feature can be observed in ZrTe$_3$, as seen in Fig. 3. Two sets of Fermi surface sheets can be extracted, as shown in Fig. 3 (a) where the red and blue lines represent the main Fermi surface  and the split Fermi surface, respectively. The corresponding band structures are shown in Fig. 3 (d) measured along k$_y$ direction for four typical momentum cuts, as marked in Fig. 3 (a). Corresponding momentum distribution curves (MDCs) at the Fermi level are plotted in Fig. 3 (c).  It is clear that, for each momentum cut, the corresponding MDCs have two sets of bands, one main band and one shoulder split band,  with a total of four peaks which can be well fitted by four Lorentzian peaks, as marked  by the four black arrows in each panel in Fig. 3(c).  The presence of four bands can be directly seen from Fig. 3(b) which represents the second derivative image from the leftmost panel of Fig. 3(d).   The detailed analysis of the band splitting around the Fermi surface gives a two-Fermi surface sheet picture shown in Fig. 3(a).  Fig. 3 (e) presents energy distribution curves (EDCs) corresponding to the measured band structure Cut 1 in Fig. 3(d).  Sharp EDC peaks are observed near the Fermi momenta, k$_L$ and k$_R$.

In addition to the band splitting on the Fermi surface, it is clear that, for the hole-like band near $\Gamma$ £¨leftmost penal in Fig. 3(d)), there is a sharp band with its top at $\sim$240 meV (see also EDCs in Fig. 3(e)), and a broad distribution of spectral weight extending to a binding energy of $\sim$120 meV.  Similar result was reported before that was suggested to be caused by the bilayer-splitting effect at the sample surface layers\cite{HBerger2009MHoesch}. Our results indicate that the feature near the binding energy of $\sim$200 meV  does not represent two well-defined bands, but one well-defined band plus an envelope of  braod spectral weight distribution. There are similar results observed in recent work of ZrTe$_5$\cite{ZhangY2017}.  Such broad shoulder and hump-like feature that appears on top of a  sharp band can be caused by k$_z$-effect\cite{ZhangY2017}.

Figure 4 shows band structure of ZrTe$_3$ with the momentum cuts perpendicular to the 3D Fermi surface which facilitates the study of band dispersion and electron dynamics. The location of the 10 momentum cuts are marked in Fig. 4(a) and the corresponding band structure are shown in Fig. 4(b).  Such a free choice of cut direction in momentum space only becomes possible because of the unique ARToF 3D-data property which maintains the same data continuity and measurement condition among the entire 2D area. For each image in Fig. 4(b), the band structure is analysed by MDC fitting with two Lorentzian peaks. The obtained band dispersion of the main band is plotted on each panel of Fig. 4(b). From these quantitative MDC fitting, the MDC width at the Fermi level  along the Fermi surface, and the Fermi velocity, can be obtained, as shown in Fig. 5. Fig. 5(a). shows the position of ten Fermi momenta along the Fermi surface corresponding to 10 momentum cuts in Fig. 4(a). The EDCs along the Fermi surface on these 10 Fermi momenta are shown in Fig. 4(b), and the corresponding MDCs at the Fermi level along the 10 momentum cuts are shown in Fig. 4(c). The EDC width, full width at half maximum, is plotted in Fig. 5(d) along the Fermi surface, which vary between 30 to 40 meV. The MDC width (full width at half maximum), on the other hand,  increases when the momentum shifts from the central region to the tip region along the 3D Fermi surface. The Fermi velocity, obtained from the extracted MDC dispersions in Fig. 4(b),  shows a slight maximum in between the central and tip regions of the 3D Fermi surface. The maximum and minimum Fermi velocities are about 4.19eV $\cdot$ \AA\ (corresponding to 6.6$\times$10$^5$ m $\cdot$ s$^{-1}$) and 2.18 eV $\cdot$ \AA\ (corresponding to 3.4$\times$10$^5$ m $\cdot$ s$^{-1}$), respectively. The extraction of these quantities in Fig. 5 will be important in understanding the role of the 3D Fermi surface in dictating the physical properties of ZrTe$_3$.

The major topic of this work is to investigate the temperature dependence of the 3D Fermi surface and its relationship with the CDW transition.  The first issue addressed here is relative change of the 3D Fermi surface topology with temperature. Due to the oval shape and crystal symmetry, the Fermi surface distance is defined in MDCs at the Fermi level Fig. 6 (b) measured along typical momentum cuts as indicated in Fig. 6 (a). The distance between the two branches of main band (black arrows) along the Fermi surface short axis or vertical direction (k$_x$ = 0, 0.05, 0.1 \AA$^{-1}$) at the Fermi level, can be used to measure the 3D Fermi surface size change and its temperature dependence is presented in Fig. 6 (d). A clear increase of the Fermi surface distance with decreasing temperature in the scale of 0.01 \AA$^{-1}$ has been revealed between 120 K and 20 K, which represents nearly 3 \% relative expansion with decreasing temperature. This change of the Fermi surface distance can be accurately measured due to the utilization of ARToF-ARPES technique. The three momentum cuts show similar variation of Fermi surface distance with temperature, as shown in Fig. 6(d). This suggests an overall expansion of the 3D Fermi surface with decreasing temperature in ZrTe$_3$. It is natural to ask whether this observation can be caused by the lattice shrinkage with decreasing temperature.  Neutron scattering measurements\cite{WTremel1998RSeshadri} of ZrTe$_3$  demonstrate a shrinkage of the lattice constant about 0.1 \% from 120 K to 20 K. Such a small change of lattice constant clearly cannot account for the $\sim$3\% change of the 3D Fermi surface size we have observed.

In order to understand the origin of the 3D Fermi surface change with temperature, we also measured the variation of the band position for the 200 meV band near $\Gamma$ at different temperatures.  To be more specific, this top-most energy location can be determined by the lorentzian-fit of the EDC spectra extracted at $\Gamma$ point, as plotted in Fig. 6 (c). The fitted temperature dependence of the peak position is shown in Fig. 6 (e). It reveals about 10 meV band shift upwards when the sample temperature changes from 120 K to 20 K.  To our best knowledge, this is the first time that such a significant band shifting with temperature is observed for the 3D-Fermi surface of ZrTe$_3$.  If we assume a rigid band shift, such a 10 meV energy shift of the chemical potential would give rise to $\sim$0.006 \AA$^{-1}$  Fermi surface distance change when the Fermi velocity is assumed to be 3 eV $\cdot$ \AA.  This is close to the observed 0.01 \AA$^{-1}$ change of the Fermi surface distance, indicating the chemical potential shift plays a major role in causing the Fermi surface distance change with temperature.   The reason behind such a considerable chemical potential shift with temperature can be either by the lattice constant variation or by strong density of states (DOS) unbalance distribution around the Fermi level, similar to the mechanism in semiconductors. It is well known in the semiconductor community that the Fermi level can be tuned away from the band gap middle. In this case, the model of Fermi level or the chemical potential is E$_F$ = E$c$ - K$_B$T$\cdot$log(N$_c$/N$_v$), where E$c$ is the energy of conduction band, K$_B$ is the Boltzmann constant, T is the absolute temperature, N$_c$ is the effective DOS in the conduction band, and N$_v$ are the effective DOS in the valance band. The E$_F$ can be tuned if N$_c$/N$_v$ is not equal to one. Recent examples from WTe$_2$\cite{Kaminski2015} and ZrTe$_5$\cite{ZhangY2017} have already shown that the Fermi level E$_F$ can change with temperature. The semi-metal character of ZrTe$_3$ suggests that the E$_F$ shift with temperature might arise from similar mechanism found in WTe$_2$ and ZrTe$_5$.

Figure 7 presents quasiparticle scattering rate change with temperature along the 3D Fermi surface in ZrTe$_3$. The temperature-dependence of the MDCs at the Fermi level and the EDCs at the Fermi momentum k$_F$  of the main band are presented in Fig. 7 (b) and (c), respectively. The black circles or lines are experimental data in both sub-figures and red lines represent corresponding Lorentzian-fit results.  As the temperature decreases from 120 K to 30 K, the corresponding fitted MDC width in Fig. 7 (d) clearly shows an non-monotonic behavior. The MDC width first decreases from 120 K to about 60 K,  suggesting a drastic decrease of the quasiparticle scattering rate.  By further lowering the temperature towards to 30 K, the MDC width rises up again implying an increase of the quasiparticle scattering rate at low temperature. The overall temperature dependence of the MDC width reveals a clear minimum that is close to the CDW coherent transition temperature T$_{CDW}$$\sim$63 K. We  applied the same analysis on all the cuts along the entire 3D Fermi surface. The fitted MDC width from several momentum cuts along the Fermi surface as function of temperature are summarized in Fig. 7(d). They all show similar behavior with a minimum located at T$_{CDW}$$\sim$63 K. This finding suggests that such MDC width anomaly across T$_{CDW}$ is a general property along the whole 3D Fermi surface. To our best knowledge, it is the first spectroscopic signature to find that the 3D Fermi surface is directly associated with the CDW order in ZrTe$_3$. Following the same procedure, one can also apply similar linewidth analysis of EDCs at k$_F$ from the main band which is associated with quasiparticle scattering rate. It reveals similar temperature-dependent behavior with the MDC linewidth, as shown in Fig. 7 (e). The quasiparticle scattering rate decreases from 120 K, reaches a minimum around  T$_{CDW}$., and rises up again at low temperature.

The consistent finding from both MDC and EDC linewidth analyses uncovers a new scattering channel of quasiparticle in the main band in the CDW state.  On the one hand, the reduction of the quasiparticle scattering rate above T$_{CDW}$ is consistent with the increase of the metallicity as observed in the resistance measurement (Fig. 1 (b)). On the other hand, its rising below T$_{CDW}$ is not compatible with the transport results. Except for the hump structure popping out along a-axis around T$_{CDW}$, the resistance along both a- and b-axes shows a metallic behavior towards low temperature. This inconsistency between the temperature evolution of the scattering rate and the resistance clearly suggest two important issues. First, the hump anomaly and further metallicity below T$_{CDW}$ in transport property can be contributed by the quasi-1D Fermi surface on the BZ boundary, which plays a dominant role in the CDW formation at low temperature. This finding also agrees with earlier works\cite{KYamaya2005TYokoya,HBerger2009MHoesch}. Second, it is quite unusual that the CDW formation impacts on the 3D Fermi surface quasi-particle properties even though it plays a minor role in generating the CDW order state in ZrTe$_3$.  In a classical CDW system, when some part of the Fermi surface is gapped out at low temperature, the rest part of the electronic states should be less coupled and scattered since some scattering channels have been blocked.  On the contrary, here the electronic states of the main band of the 3D Fermi surface suffer extra scattering process while part of the quasi-1D Fermi surface has been gapped out. Especially if one stays with the conventional nesting picture for CDW in ZrTe$_3$, it would be quite unusual to find such an involvement of the electronic states in the CDW state which is far from the nesting part of the Fermi surface.

To achieve more insight on understanding such an electronic renormalization effect developed in the CDW state, self-energy analysis of the measured ARPES spectral function in semi-quantitative level has been applied on the very same data based on the single-particle Green's function model, as shown in Fig. 8.  The measurement temperature is 30 K which is below T$_{CDW}$. The E-k image (Cut1) is shown in Fig. 8(a) and the fitted dispersions are shown in Fig. 8(b) measured along several typical momentum cuts as indicated in Fig. 7 (a). Here the so-called bare band is approximated as a single straight line linking between the k$_F$ and the band location at E$_b$ = 100 meV. The MDCs  at each binding energy are fitted by the Green's function, from which the real and imaginary parts of the self-energy can be extracted, as shown in Fig. 8 (c) and  (d), respectively. The overall self-energy is small; it remains almost unchanged except in the low energy region  where a clear kink appears in the real part of self-energy at an energy of  $\sim$20 meV.   For a metallic system like ZrTe$_3$, the electron-phonon coupling can serve as the most natural explanation for the self-energy anomaly observed here. This means the low energy phonon, especially those with 20 meV in energy scale, can couple with the main band electronic states in the CDW order. However, we note that the overall electron-phonon coupling is weak.  At present, there is no straight answer to the question on how the CDW formation is related to the electron-phonon coupling along the 3D Fermi surface.


In summary, we have performed high resolution laser-based ARPES measurements on ZrTe$_3$ and focused on studying the 3D-hole-like Fermi surface around $\Gamma$ and related electron dynamics.  We have clearly resolved double Fermi surface sheets for the 3D hole-like Fermi surface around the zone center. The 3D Fermi surface shows a pronounced shrinkage with increasing temperature.  In particular, the quasiparticle scattering rate along the 3D Fermi surface experiences an anomaly near the charge density wave transition temperature of ZrTe$_3$ ($\sim$63 K).  Signature of electron-phonon coupling is observed with a dispersion kink at $\sim$20 meV; the strength of the electron-phonon coupling around the 3D Fermi surface is rather weak. These results indicate that the 3D Fermi surface is  closely coupled to the charge-density-wave transition  and  suggest a more global impact on the entire electronic structure induced by CDW phase transition in ZrTe$_3$.

\vspace{3mm}

\noindent {\bf Acknowledgement}\\
 Project supported by the National Basic Research Program of China (Grant No. 2015CB921301), the National Natural Science Foundation of China (Grant Nos. 11574360, 11534007, and 11334010), and the Strategic Priority Research Program (B) of the Chinese Academy of Sciences (Grant No. XDB07020300). 

\noindent{\bf Additional information}\\
Correspondence and requests for materials should be addressed to L. Y.(li.yu@iphy.ac.cn) and X.J.Z.(XJZhou@iphy.ac.cn)

\newpage

\begin{figure*}[tbp]
\begin{center}
\includegraphics[width=1\columnwidth,angle=0]{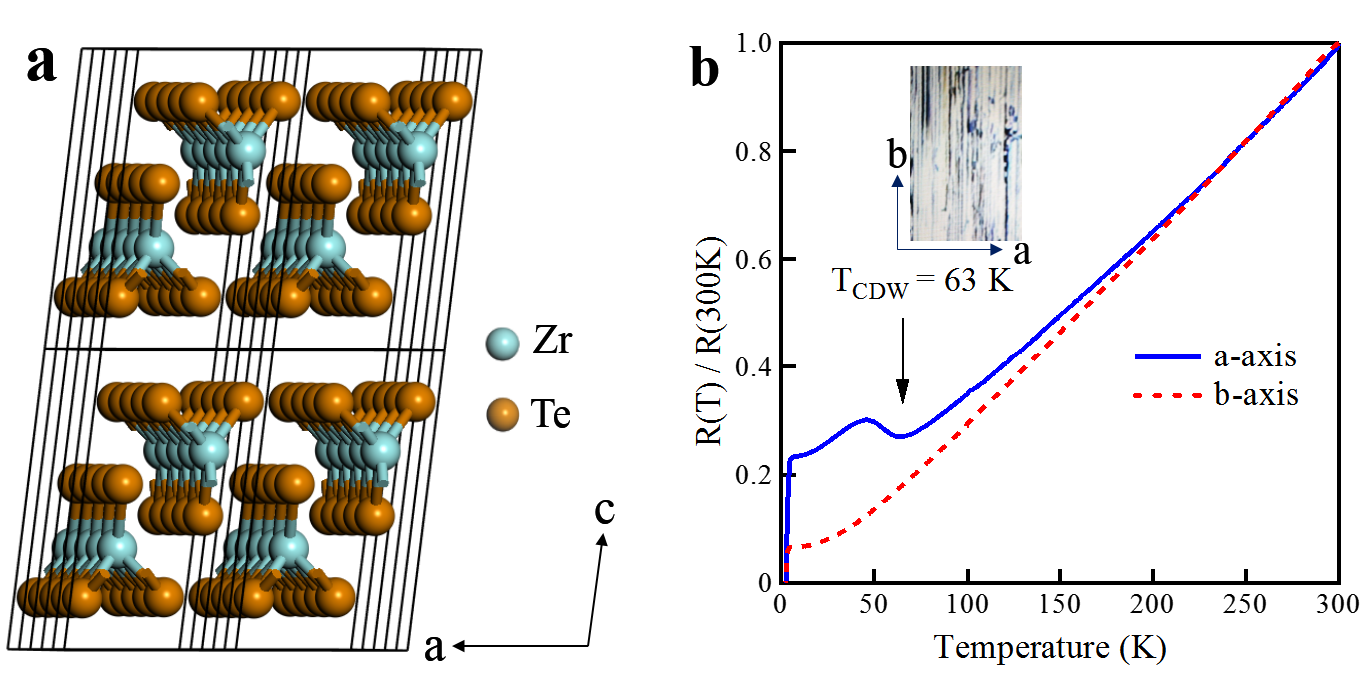}
\end{center}
\caption{{\bf Crystal structure and temperature dependence of resistivity of ZrTe$_3$.} (a) Crystal structure of ZrTe$_3$. The green spheres represent Zr atoms and the yellow ones represent Te atoms. (b) Temperature dependence of normalized resistivity for our ZrTe$_3$ single crystal samples.  There is a clear resistivity hump at $\sim$ 63 K along the a-axis and a superconducting transtion at $\sim$ 2 K along both a- and b-axes. The cleaved surface morphology of a ZrTe$_3$ sample is shown in the inset of  (b); there are one-dimensional thread-like structures running along the b-axis. }
\end{figure*}

\begin{figure*}[tbp]
\begin{center}
\includegraphics[width=1\columnwidth,angle=0]{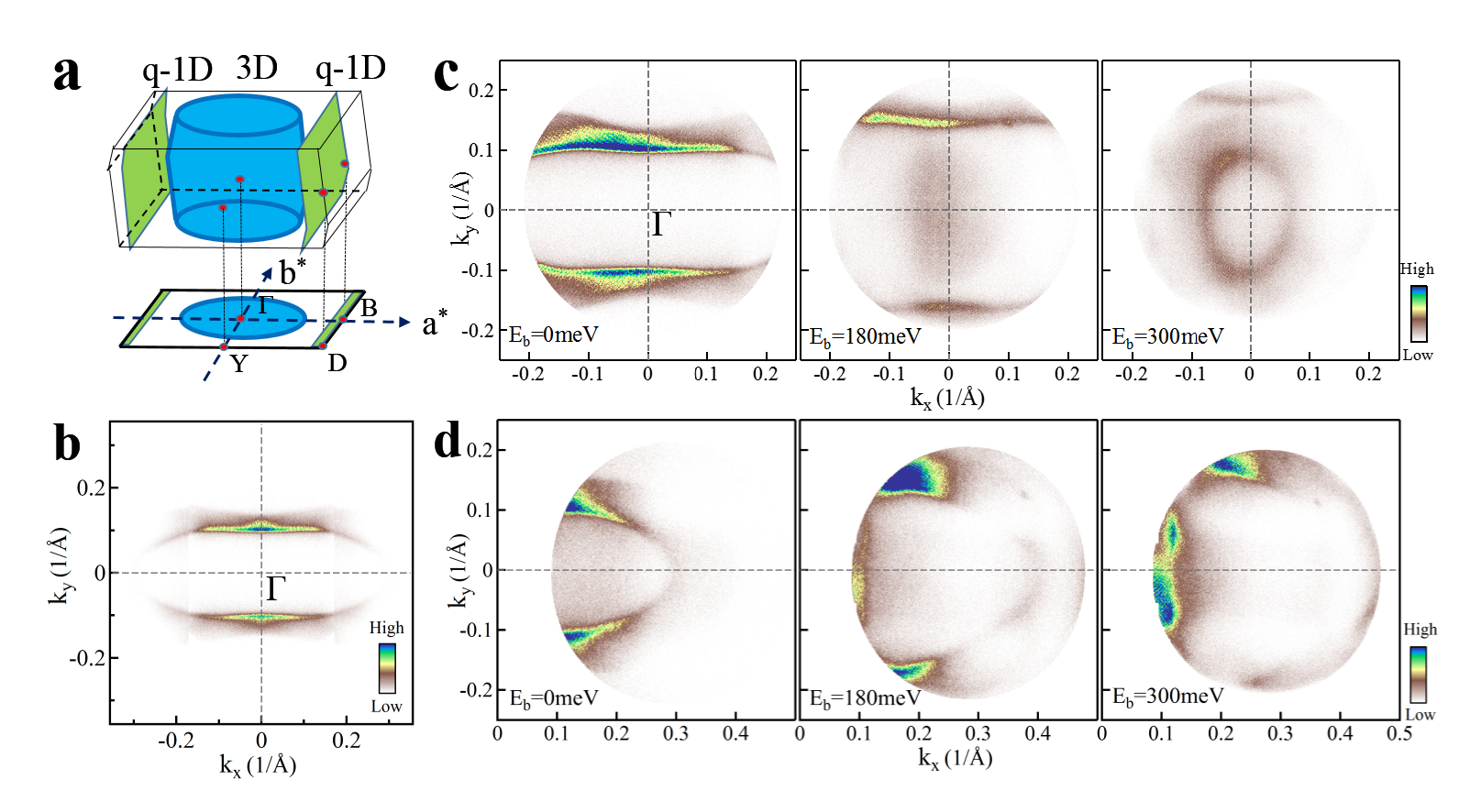}
\end{center}
\caption{{\bf Fermi surface and constant energy contours of ZrTe$_3$.} (a) Schematic of  ZrTe$_3$ Brillouin zone. The upper one is schematic diagram of the first Brillouin Zone and two sheets of Fermi surface. The central cylindrical shape is the three-dimensional  Fermi surface. The  quasi-one dimensional Fermi surface is near the boundary of the first Brillouin zone.  Projection of the first Brillouin zone and the Fermi surface along ab plane is shown below the 3D Brillouin zone. High-symmetry points are indicated. (b) The complete 3D Fermi surface which is jointed and symmetrized from two measurements. (c, d) Constant energy contours of ZrTe$_3$ of two measurements at different binding energies of 0, 180 and 300 meV, respectively. The spectral intensity is integrated within 10 meV with respect to each binding energy. The measurement geometry is set under s polarization. }
\end{figure*}

\begin{figure*}[tbp]
\begin{center}
\includegraphics[width=1\columnwidth,angle=0]{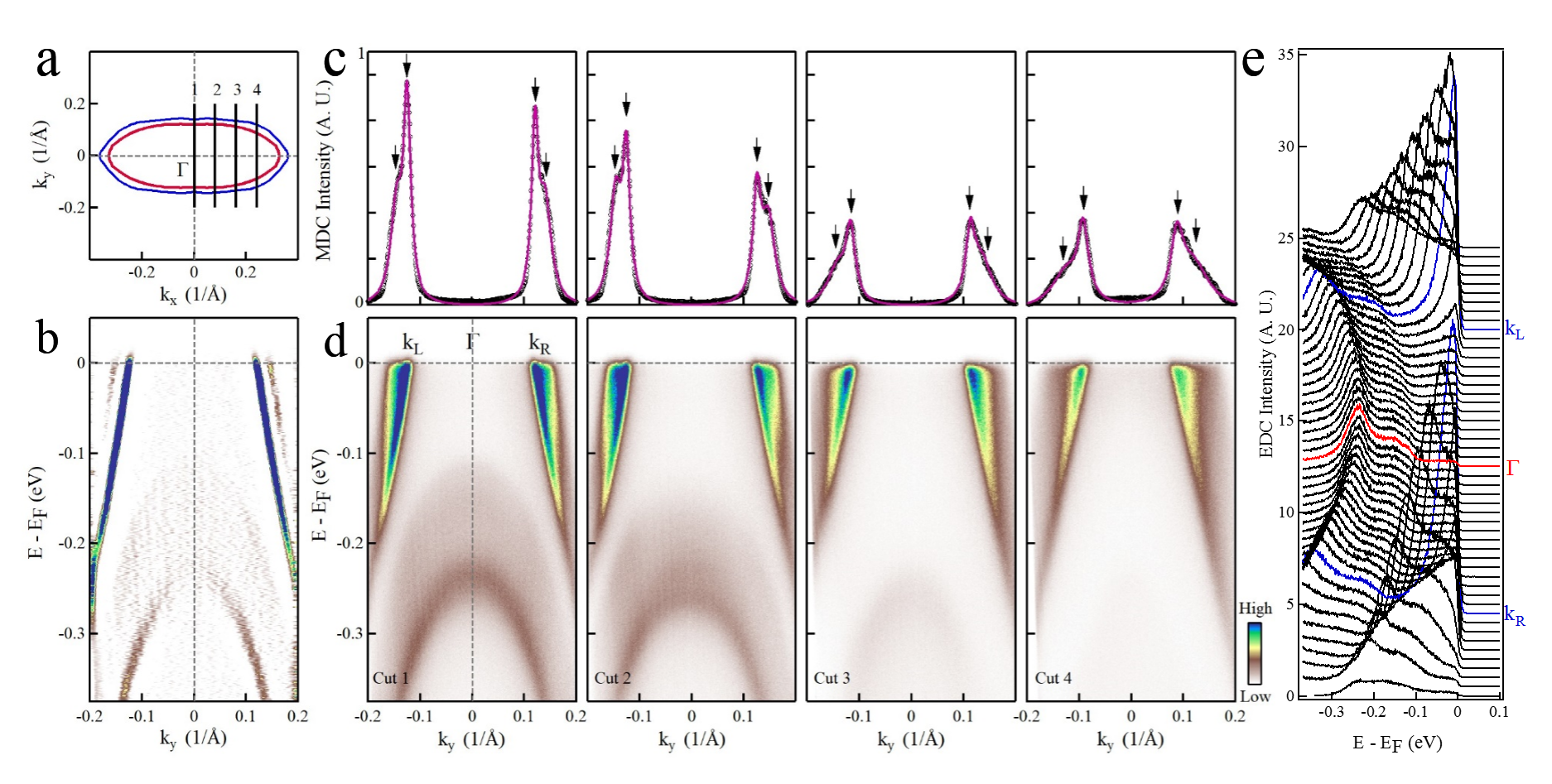}
\end{center}
\caption{{\bf Observation of splitting features in ZrTe$_3$.} (a) Fitted Fermi surface of ZrTe$_3$ that has two Fermi surface sheets. Red and blue lines represent main sheet and split sheet, respectively. Band structures measured along momentum Cuts 1, 2, 3 and 4 are shown in (d). The locations of these momentum cuts are indicated in (a). Corresponding momentum distribution curves (MDCs) at the Fermi level for band structures in (d) are shown in (c). Four peak positions are marked by the arrows in (c). (b) MDC 2nd derivative band structure of Cut 1. (e) Energy distribution curves (EDCs) of Cut 1. EDCs of $\Gamma$ point and k$_F$ are indicated by red line and blue line, respectively. }
\end{figure*}

\begin{figure*}[tbp]
\begin{center}
\includegraphics[width=1\columnwidth,angle=0]{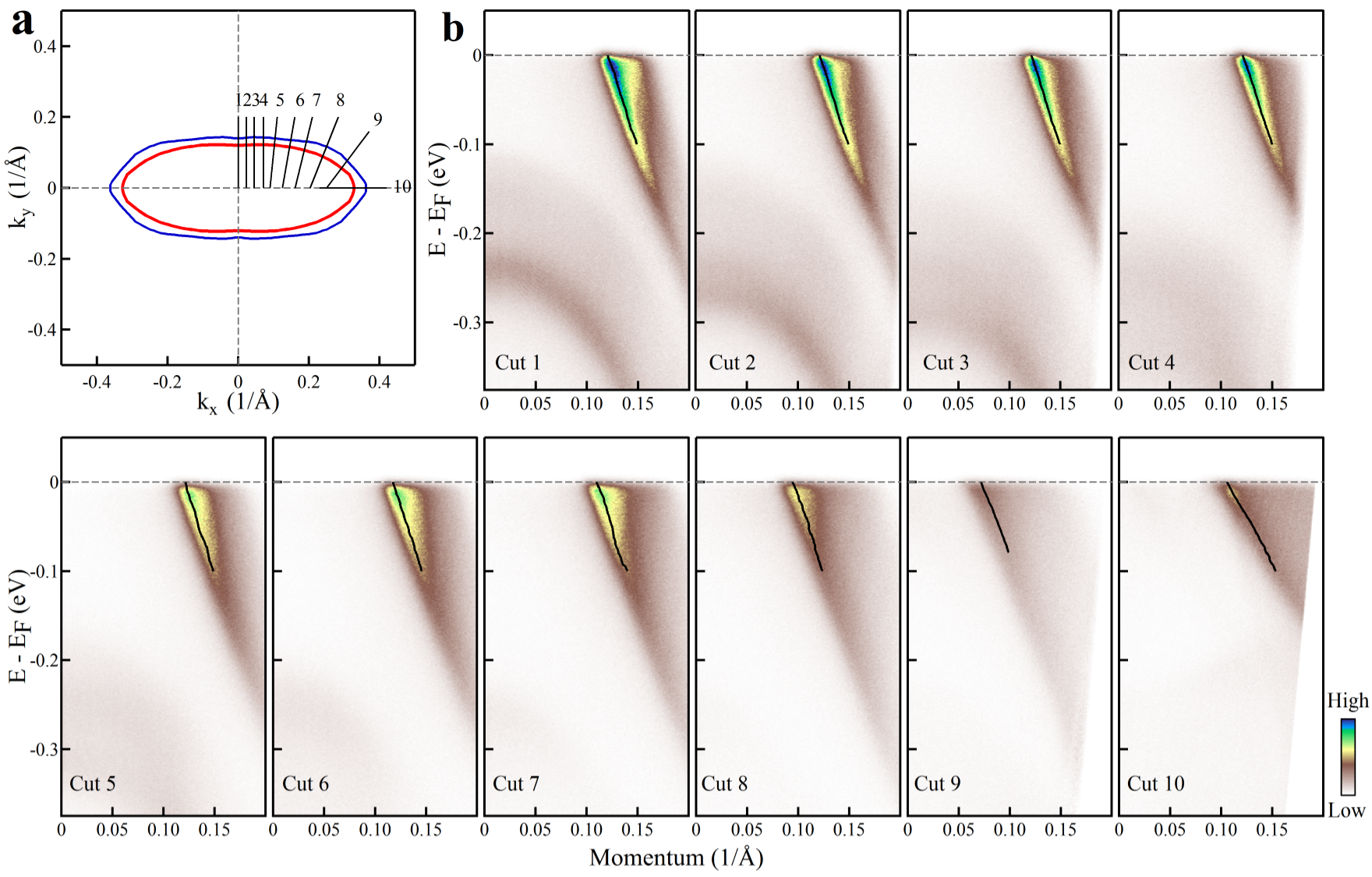}
\end{center}
\caption{{\bf Band structure of ZrTe$_3$ with momentum cuts perpendicular to the Fermi surface.} (a) Fitted Fermi surface lines by two lorentz functions. Red and blue lines represent main Fermi surface and split Fermi surface, respectively. (b) Band structures measured along momentum cuts 1 to 10. The locations of these momentum cuts are indicated in (a). Fitted dispersions of main band by Lorentz functions are shown in (b), and indicated by black lines. }
\end{figure*}

\begin{figure*}[tbp]
\begin{center}
\includegraphics[width=0.8\columnwidth,angle=0]{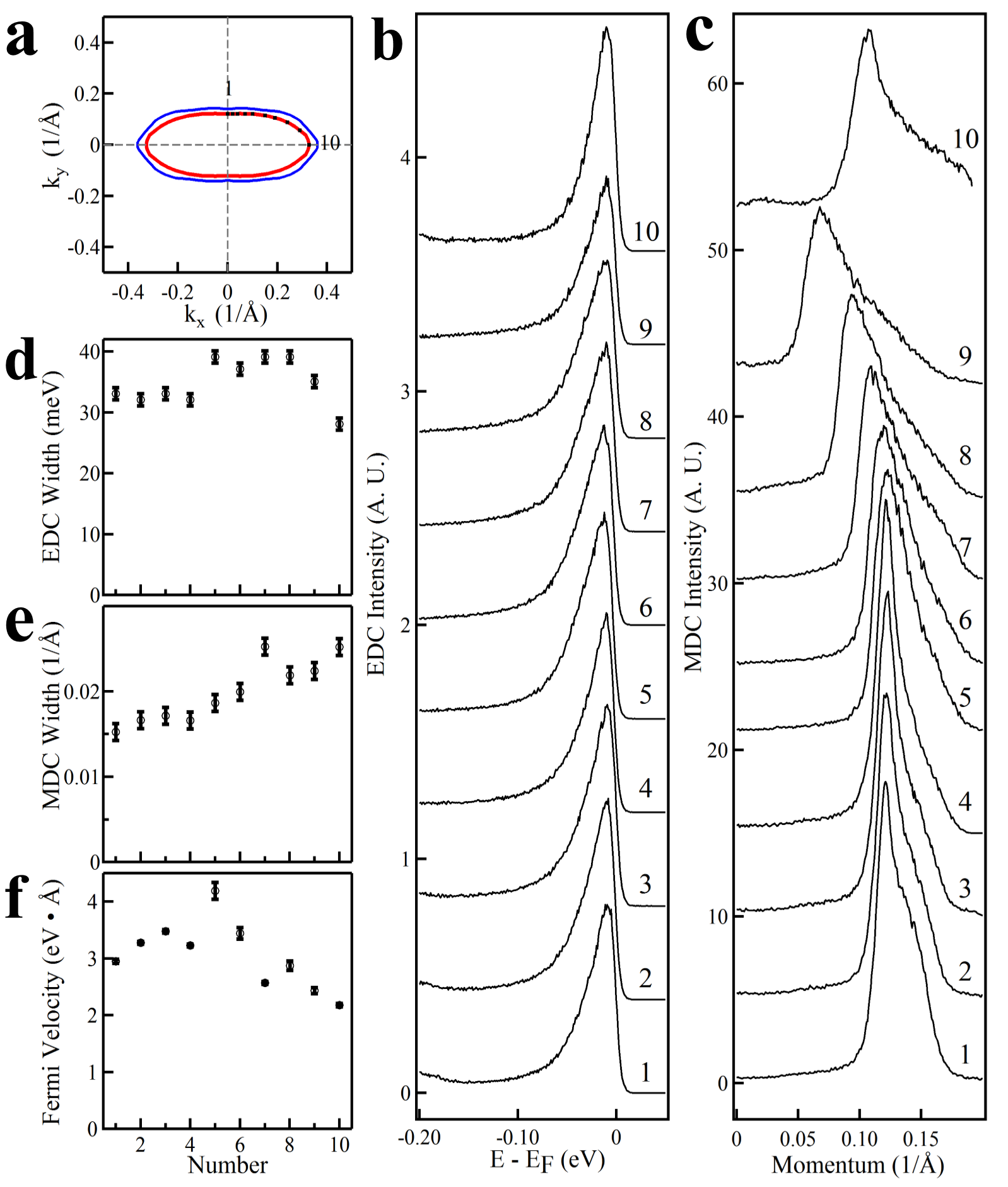}
\end{center}
\caption{{\bf Scattering rate and Fermi velocity of ZrTe$_3$ along the 3D Fermi surface.} (a) Fitted Fermi surface lines by two lorentz functions. EDCs and MDCs along momentum points 1 to 10 are shown in (b) and (c), respectively. The locations of these momentum points are indicated in (a), which are acquired from Fig. 4 (b). (d) and (e) indicate EDC width and MDC width, respectively. (f) Evolution of Fermi velocity along the Fermi surface, which are acquired from fitted dispersions of Fig. 4 (b). }
\end{figure*}

\begin{figure*}[tbp]
\begin{center}
\includegraphics[width=1\columnwidth,angle=0]{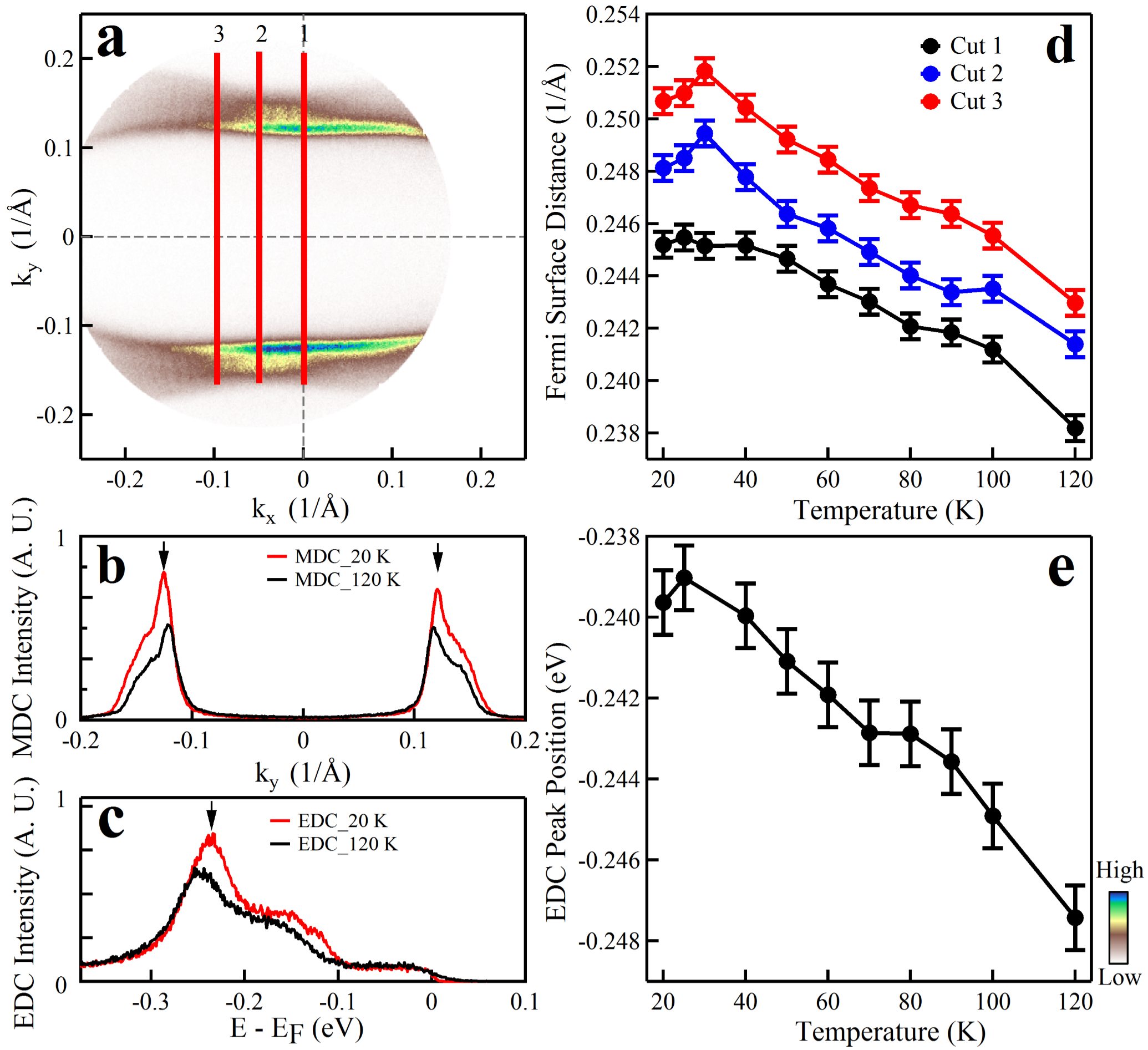}
\end{center}
\caption{{\bf Observation of 3D Fermi surface shrinkage with temperature in ZrTe$_3$.} (a) Fermi surface acquired by one time measurement. (b) MDCs at the Fermi level of Cut1 at different temperatures.  The arrows represent the peak position, and the distance between two peaks is marked as Fermi surface distance (FSD). Evolution of FSD with temperature are shown in (d), the locations of these momentum cuts are indicated in (a). (c) EDCs at $\Gamma$ point at different temperature. The arrow represents the EDC peak position. Evolution of EDC peak position with temperature is shown in (e). }
\end{figure*}

\begin{figure*}[tbp]
\begin{center}
\includegraphics[width=1\columnwidth,angle=0]{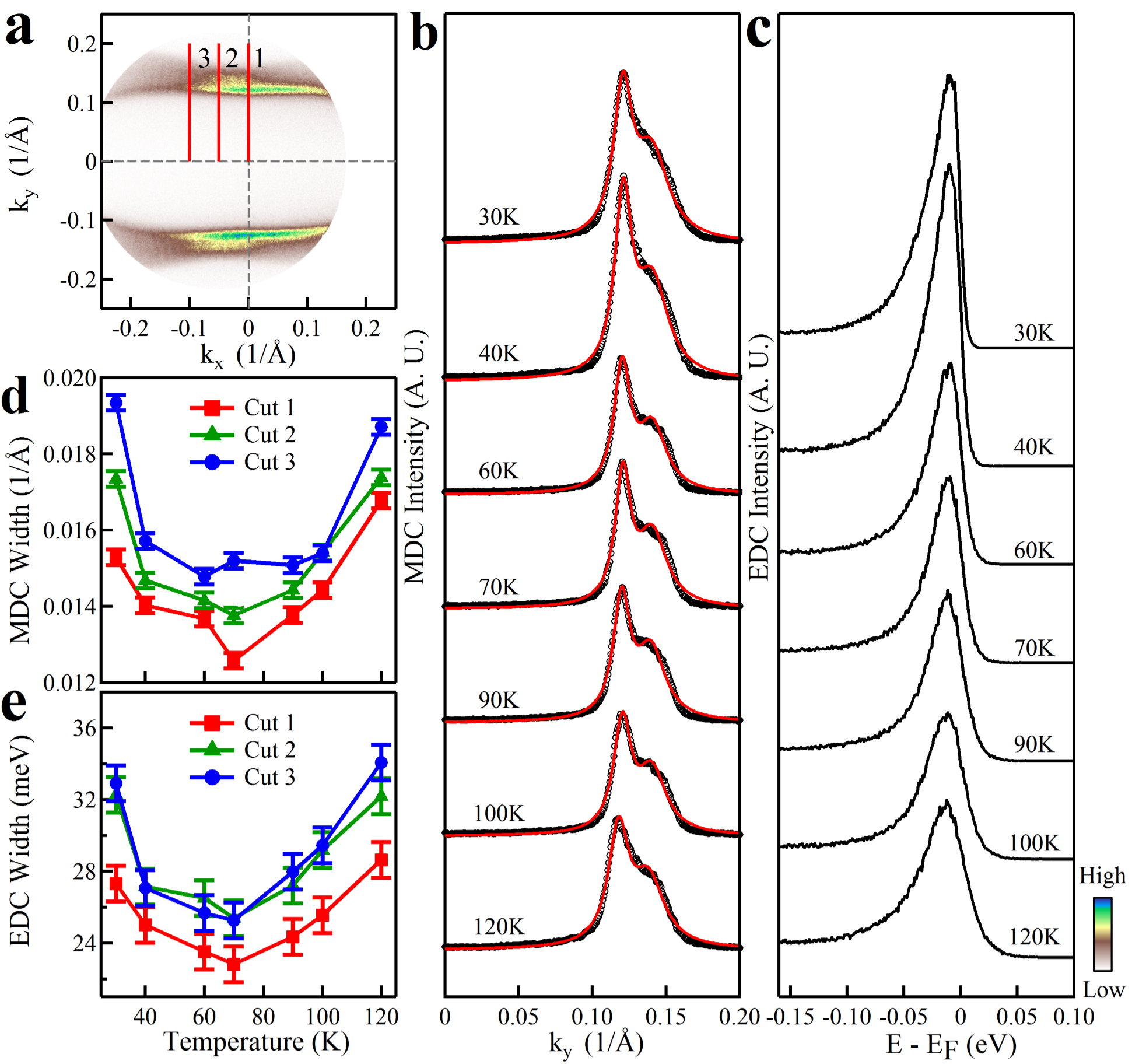}
\end{center}
\caption{{\bf Scattering rate of ZrTe$_3$ at different temperatures.} (a) Fermi surface acquired by one time measurement. (b) MDCs at Fermi level of Cut1 at different temperatures. Black circles and red lines represent the raw data and fitted results by lorentz functions, respectively. Evolution of MDC width with temperature is shown in (d). (c) EDCs at k$_F$ of main band at different temperatures. Evolution of EDC width with temperature is shown in (e). The locations of these momentum cuts are indicated in (a). }
\end{figure*}

\begin{figure*}[tbp]
\begin{center}
\includegraphics[width=1\columnwidth,angle=0]{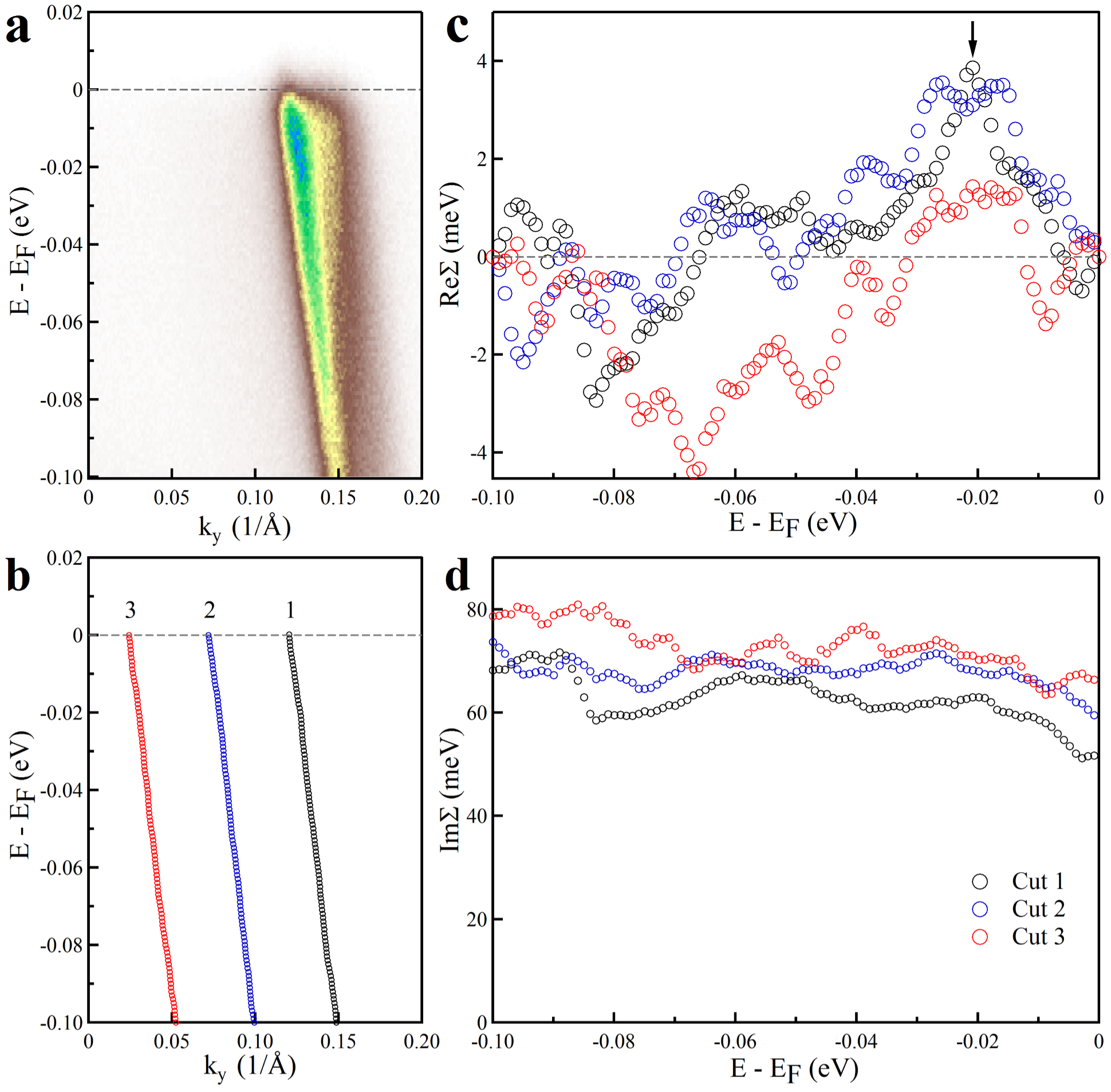}
\end{center}
\caption{{\bf Self-energy of ZrTe$_3$ measured at T = 30 K.} (a) Band structure of Cut1 indicated in Fig. 7 (a). (b) Band dispersions in a small energy window from E$_b$ = 0 to E$_b$ = 100 meV acquired from MDC fitting. Black, blue and red circles represent the cut 1, cut 2 and cut 3, respectively. For clarity, cut 2 and cut 3 are offset by 0.05 and 0.1 \AA$^{-1}$, respectively. (c) The effective real part of the electron self-energy for the 3 cuts, which are indicated in Fig. 7 (a). The bare bands are chosen for each band as straight lines connecting the two points on the measured dispersion at the Fermi level and 100 meV binding energy. The arrow indicates the kink position. (d) The imaginary part of the electron self-energy for the 3 cuts. }
\end{figure*}





\end{document}